\documentclass[runningheads]{llncs}
\usepackage{graphicx}
\usepackage{verbatim}
\usepackage{booktabs} % For formal tables
\usepackage{enumitem}
\usepackage{pgfplots}
\pgfplotsset{
	legend image code/.code={
		\draw [#1] (0cm,-0.1cm) rectangle (0.6cm,0.1cm);
	},
}

\begin{document}
\title{CBPF: leveraging context and content information
	for better recommendations}
\titlerunning{CBPF}
\author{Zahra Vahidi Ferdousi%\inst{1}\orcidID{0000-1111-2222-3333} 
	\and Dario Colazzo%\inst{2,3}\orcidID{1111-2222-3333-4444} 
	\and Elsa Negre%\inst{3}\orcidID{2222--3333-4444-5555}
}
\authorrunning{Z. Vahidi Ferdousi et al.}
\institute{Paris-Dauphine University, PSL Research University, CNRS UMR 7243,\\ LAMSADE, Paris, France\\
\email{\{zahra.vahidiferdousi, dario.colazzo, elsa.negre\}@dauhphine.fr}}
\maketitle              % typeset the header of the contribution
\begin{abstract}
Recommender systems help users to find their appropriate items among large volumes of information. Different types of recommender systems have been proposed. Among these, context-aware recommender systems aim at  personalizing as much as possible the recommendations based on the context situation in which the user is. In this paper we present an approach integrating contextual information into the recommendation process by modeling either item-based or user-based influence of the context on ratings, using the Pearson Correlation Coefficient. The proposed solution aims at taking advantage of content and contextual information in the recommendation process. We evaluate and show effectiveness of our approach on three different contextual datasets and analyze the performances of the variants of our approach based on the characteristics of these datasets, especially the sparsity level of the input data and amount of available information.

\end{abstract}

\section{Introduction}

Nowadays, we are faced with a rise of the amount of data on the web, provided by different sources. As a consequence, a user can quickly be overwhelmed by the huge volume of information. \textit{Recommender systems (RS)}~\cite{jannach2010recommender} aim to help the user to find her appropriate information among all others. Recommendations are principally based on two main approaches: \textit{content-based} and \textit{collaborative filtering}. In the former, characteristics  of items/users are used to find and recommend similar items to the ones the user liked in the past. In the latter approach, similar users are found based on the previous users' preferences, then items that these similar users liked in the past are recommended.
These traditional recommender systems have proved their effectiveness in different areas~\cite{ricci2011introduction}, including music, movies, places of interest, news, research articles, online courses, etc. But they have the limitation of not considering the contextual situation in which the user is, at the moment she wants to use the item. In fact this information can roughly influence her preferences for items~\cite{adomavicius2011context}. As an example, when choosing a movie to watch, the user will have different preferences depending on whether she wants to watch the movie with a kid or with her partner. In this case, a context-aware recommender system (CARS), integrating such contextual information about the user in the recommendation process, can provide more relevant recommendations~\cite{adomavicius2005incorporating}. \\
A particular class of CARSs are based on \emph{pre-filtering}, based on the idea of pre-procesing contextual data so as to tune the input of  a given (traditional) RS in order to increase its effectiveness. Along the lines of a preliminary investigation presented in the workshop paper~\cite{vahidi2018cbpf}, where we proposed a pre-filtering CARS that integrates contextual information about users by modeling them with item-based influence of context on ratings, in this paper we propose the user-based version of this approach, and we present here results on a much more extensive experimental analysis. With respect to CARS state of the art (discussed later on) our approach, named Correlation-Based Pre-Filtering is, in a sense, more user-centric, as  we propose to model item/user-based influence based on the item- or user-based Pearson Correlation Coefficient (PCC)~\cite{benesty2009pearson} between context and ratings. The distinctive feature of using PCC allows us to catch more precisely the influence of context on ratings, and so to compute more accurate similarities between contexts, which is a crucial point in our pre-filtering process. In addition, we  use content information about items/users to improve our model, like, for instance, the category of a film or the age or gender of users. %\DC{This last sentence seems incomplete: what cahracteristics? How?}
Our experimental analysis on three typically used datasets show improvements over state of the art approaches.\\
%\DC{changed a bit above}
%\DC{I think that we could leave the above as it is, while the paragraph below details new contribution; we should be safe, but this also depends from each single reviewer, even if in general extended version of ws paper appear in conferences, in a consolidated version, as for our case} 
%\DC{Please double check}
With respect to our preliminary investigation presented in the workshop paper ~\cite{vahidi2018cbpf}, in this paper our new contributions are the followings: we propose to model the context by relying on the user-based influence of contexts on ratings; we compare the item- and user-based approaches, and study the cases where each one of these versions can perform the best. In our experimental analysis we use three different datasets from three different domains to highlight previously unseen properties. And we demonstrate that our approach can deal well with either sparse or dense data.\\

The remainder of the paper is organized as follows: in the next section we present a state of the art of the subject. In Section \ref{proposition} we describe our approach. In Sections \ref{expSetup} and ~\ref{results} we respectively describe the  setup and results of our experimental analysis.  Finally we discuss our results and make conclusive remarks. 

\section{Related Work}\label{relatedWork}

%\DC{Please double check} 
CARSs aim to take into account the users' contextual information, in the most efficient way, in order to propose more relevant and personalized recommendations~\cite{adomavicius2011context}. So instead of the 2D rating function of traditional RSs ($R: user \times item \rightarrow rating$), in CARSs we  have the multidimensional function, $R: user \times item \times context \rightarrow rating$~\cite{adomavicius2005incorporating}. The context of a user is composed of a number of context factors like \textit{time, location, weather, companion, etc}. To each one of these context factors some values can be associated, called context conditions. For example possible context conditions for \textit{time} could be \textit{morning, afternoon, evening} and \textit{night}, and for \textit{companion} could be \textit{alone, friend, family, etc}.\\
The integration of contextual information in CARSs can be done by relying on either \textit{pre-filtering}, \textit{post-filtering} or \textit{contextual modeling}~\cite{adomavicius2011context}:\vspace{0.1cm}

In \textbf{pre-filtering} approaches, contextual information is used to select only appropriate data for the target user's context situation, and then a traditional recommendation technique is applied on this selection.
Numerous approaches have been proposed in this category. We can mention: the \textit{reduction-based} approach~\cite{adomavicius2005incorporating}, with its two variants \textit{exact pre-filtering} and \textit{generalized pre-filtering}; the splitting approaches: \textit{item-splitting}~\cite{baltrunas2014experimental}, \textit{user-splitting}~\cite{baltrunas2009towards} and \textit{UI-splitting}~\cite{zheng2014splitting}; the \textit{differential context modeling} approach, and its two variants \textit{differential context relaxation} (DCR) and \textit{differential context weighting} (DCW), proposed in \cite{zheng2012optimal}; and the \textit{distributional semantic pre-filtering} approach~\cite{codina2016distributional}.\vspace{0.1cm}

\textbf{Contextual modeling} approaches try to extend traditional recommendation techniques by integrating directly contextual information into the recommendation algorithm. Some of the most popular propositions in this category are the following: \textit{Tensor factorization (TF)} and its variants \textit{multiverse recommendation} ~\cite{karatzoglou2010multiverse} and \textit{factorization machine} ~\cite{rendle2011fast}; \textit{the deviation-based context-aware matrix factorization (CAMF)}~\cite{baltrunas2011matrix} with its several derived model: \textit{CAMF-C}, \textit{CAMF-CI}, \textit{CAMF-CC} and \textit{CAMF-CU}; \textit{contextual sparse linear method (CSLIM)} ~\cite{zheng2014cslim}; \textit{the similarity-based approaches of CAMF and CSLIM}~\cite{zheng2015similarity} with their three versions \textit{ICS}, \textit{LCS} and \textit{MCS}; and the \textit{context-aware collaborative filtering} proposed by~\cite{chen2005context}.\vspace{0.1cm}

In \textbf{post-filtering} approaches, first a context-free recommendation algorithm is applied on the data and then the resulting recommendation list is contextualized by filtering or reordering items.
This category of approach has received less attention than the two previous categories, but we can still cite the \textit{weight post-filtering} and \textit{filter post-filtering} approaches proposed by ~\cite{panniello2009experimental}; and the \textit{content-based post-filtering} model ~\cite{hayes2004context}.\\

Among these previous approaches, we are especially interested into the following approaches: \textit{DSPF}, \textit{deviation-based and similarity based CAMF} and \textit{DCM}, which are the most similar to our approach. In fact, they try to model the influence of the context on their model, but based on different points of view:
\textit{DSPF}~\cite{codina2016distributional} models the influence of context on ratings based on the difference between context-free rating and the rating given in the specific context. But, differently from our technique,  this influence computation is not user-centric enough because of the way the context-free rating is estimated~\cite{vahidi2018cbpf}.\\
In fact they have represented each context condition ($c$) by a vector ($w_c$)
	containing  the item-based or user-based influence of the context condition in ratings. For example in the item-based case, the influence of the context condition $c$ in ratings of item $i$, noted as  $w_{ci}$, is computed based on the difference between the rating done by a user $u$ for item $i$ in this context situation, $r_{uic}$, and the predicted context-free rating, $\hat{r}_{ui}$, as follows.
	\begin{equation}\label{eq_codina}
	w_{ci}=\frac{1}{|R_{ic}|+\beta} \sum_{r_{uic} \in R_{ic}}{(r_{uic} - \hat{r}_{ui})}
	\end{equation}

	\noindent where $R_{ic}$ is the set of ratings for item $i$ in condition $c$, and $\beta$ is a decay factor for decreasing the estimated deviation when $|R_{ic}|$ is small. This context-free rating, $\hat{r}_{ui}$, was calculated by the baseline context-free predictor of \cite{koren2015advances}, which is the sum of the overall average ratings ($\mu$) and the observed deviations of user \textit{u} ($b_u$) and item \textit{i} ($b_i$): $\hat{r}_{ui} = \mu + b_u + b_i$.
	
	After computing the representation of each context condition ($w_{c}$ = [$w_{ci_1}$ $w_{ci_2}$ $w_{ci_3}$ ...]), the contextual situation representation ($w_s$) was made by an aggregation of the context conditions which composed this context (equation \ref{eq_sum}).	\begin{equation}\label{eq_sum}
	w_s=\frac{1}{|C|} \sum_{c \in s}{w_{c}}
	\end{equation}
	%\DC{I am a bit lost in the above part, probably an example would help in explaining the mathematical model...also, please check as I changed a bit}
	In this proposition, the basic idea of representing the context by computing the item-based or user-based influence of contexts on ratings is effective, but the computation of the context-free ratings ($\hat{r}_{ui} = \mu + b_u + b_i$) is only behavior-based and not personalized. Imagine we want to estimate the context-free ratings of an over-rated science-fiction movie by two users who  have radical different interests (one who loves science-fiction movies and the other one who hates them). In this case, contrary to what is expected, the estimated ratings will be the same for these two persons. Therefore the influence calculated by this measure would be biased, and not user-centric enough.\\
\textit{CAMF}~\cite{baltrunas2011matrix} is an extension of matrix factorization~\cite{koren2009matrix}. The \textit{deviation-based} version tries to take into account the context situation of users by integrating additional model parameters in the matrix factorization equation. And the \textit{similarity-based} version integrates a similarity function that estimates the similarity between a contextual situation and a non-contextual situation.
These CAMF approaches proved their effectiveness to improve recommendation performance in comparison to context-free recommendation and some of the context-aware recommendation approaches, but like other contextual modeling approaches, and differently from ours, they have the disadvantage of needing to be implemented from scratch, with no possibility of re-using recommendation techniques already in production. \\
\textit{DCM}~\cite{zheng2012optimal} is based on the user-based collaborative filtering algorithm~\cite{su2009survey}. The authors propose to separate the algorithm into different functional components, and apply differential context constraints to each component, in order to maximize the performance of the whole
algorithm. %\EN{Differently, our approach is more user-centric, more easy to use (contrary to contextual modeling which force to re-implement) and models the context from a different point of view.}
Differently, our approach try to model the context from a different point of view. \\

\section{Methodology}\label{proposition}

In this paper we propose a new pre-filtering approach, based on the influence of context on ratings, by modeling it based on the user-based correlation between context and ratings, computed by the Pearson Correlation Coefficient.
%This approach was evaluated on a contextual dataset and the ability of this approach for extracting the influence of context on ratings based on items was proved. In this paper we want to analyze this influence based on users. We go further with an empirical study of this influence on different datasets of different domains, with a comparison between the item-based and user-based influences.\\

%\DC{The above paragraph could be a bit redundant if we keep the new paragraph in the intro; also, in this paper we do both user an product based influence right? The paragraph seems to say that we do not redo product based. }

A recommendation problem is often viewed as a matrix/tensor completion problem. A recommender system will firstly estimate missing ratings, and then it will recommend to each user her corresponding items with higher estimated rates.
In the case of pre-filtering CARSs, we want to integrate the contextual information into the estimation phase of missing ratings.
Our correlation-based pre-filtering approach, like the reduction-based pre-filtering approach~\cite{adomavicius2005incorporating}, makes the hypothesis that a user will rate an item similarly in two similar contexts.
Based on this hypothesis, to recommend an item to a user in a specific context, we can identify ratings given in similar contexts of this specific context, and apply a traditional 2D recommendation technique on this selection.
The whole recommendation process can be decomposed in five steps:\\

\noindent \textbf{Step 1 : }To be able to find similar contexts, we need a strong representation of context. In~\cite{vahidi2018cbpf} we proposed to represent contexts based on their item-based influence on ratings. In fact, the context can influence the ratings differently, according to items. For example in the case of points of interest recommendation, a snowy weather will have a positive influence on some winter sport centers, but a negative influence on natural parks. This is why it is important to compute this influence according to items.

In this paper we propose to represent contexts based on their user-based influence on ratings. Indeed, we can say that the influence of context on ratings also depends on users, and will differ from one user to another. For example, a "family person" could like to practise activities with her family, whereas another person may not like this and prefer to practise activities with her friends. So the social context will influence differently these two persons.

We can compute this influence by calculating the Pearson Correlation Coefficient (PCC) of the rating variable $r$, and each context condition variable $c_j$, with $j \in \left[1, n\right]$, where $n$ is the total number of context condition variables. We choose this correlation measure because in statistics, PCC is widely used to measure the strength of linear association between two variables, and this corresponds to what we want, since we want to catch the influence of context conditions on ratings.\\ %\DC{slightly changed here}
In a context-aware environment, an observation will be the cross-tabulation of the variables of user, item, rating and $m$ different context factors (e.g. \textit{daytype, season, location, social, etc}). To apply PCC, we transform context factors into binary variables. So let us denote with $X_t = (u_t, i_t, r_t, c_{1t}, c_{2t}, ..., c_{nt})$ the t\textsuperscript{th} observation, which represents the evaluation $r_t$ of the user $u_t$ for the item $i_t$ in the context situation $c_{1t}, c_{2t}, ..., c_{nt}$, where as said before, $n$ is the total number of context conditions, and $c_{mt}=1$ means that the m-th context condition is present in the context of the user, and $c_{mt}=0$ means that it is not present. For instance, in a movie RS with a notation from 1 to 5 stars, $X_1$= \textit{(John, Star Wars, 4, weekend=1, workingday=0, holiday=0, summer=1, winter=0, spring=0, autumn=0, home=1, public\_place=0, friend's home=0, alone=1, partner=0, friends=0)} means that \textit{John} had evaluated the movie \textit{Star Wars} by \textit{4} stars, when he watched the movie \textit{alone}, at \textit{home} in a \textit{weekend} of \textit{summer}.
\\
So the user-based correlation between the rating $r$ and a context condition $c_j$ is calculated as follows in Equation \ref{eq_pcc_user}.
\begin{equation}\label{eq_pcc_user}
w_{c_ju}=PCC_{u}(r, c_j) = \frac{\sum_{k \in K} (r_{k}- \overline{r_{u}})(c_{jk} - \overline{c_{u}})}{\sqrt{\sum_{k \in K}  (r_{k} - \overline{r_{u}})^2} \sqrt{\sum_{k \in K}  (c_{jk} - \overline{c_{u}})^2}}
\end{equation}
\noindent where \textit{K} is the set of observations $X_k = (u, i_k, r_k, c_{1k}, c_{2k}, ..., c_{nk})$ with  user $u$,  $\overline{r_{u}}$ is the mean of the ratings given by the user $u$, while  $\overline{c_{u}}$ is the mean value of the context condition $c$ over observations for user  $u$.

%\DC{Normally we should explain this equation immediately after, by describing the meaning of each variable. Zahra would it be possible to re-arrange things quickly? }\ZVF{Is it ok now?}

 Based on the above explanations, we can build a vector representation for each context condition. The size of this vector will be the total number of users, and the values (between -1 and 1) of this vector are equal to the user-based PCC between the rating vector and the binary context condition vector.\\
In real world recommendation problems, the total number of users is often very large, and the correlation calculation would be computationally consuming. To overcome this computational cost we propose to cluster users into a limited number of groups, and to compute the influence based on clusters of users. This clustering could be done based on the available static information about  users' characteristics (e.g. age, sex, etc), or directly based on the ratings.

\begin{figure}[t]
	\centerline{\includegraphics[width=6cm]{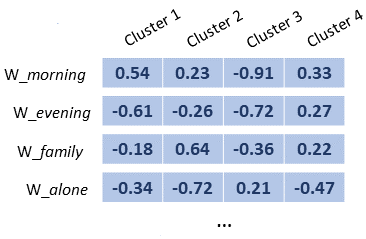}}
	\caption{Examples of representation of cluster-based context condition (Step 1)}
	\label{fig_cc}
\end{figure}
%\DC{in figure 3 you could put W\_<.... in the caption so to make room for the part about clusters and make the figure less tiny}
Figure~\ref{fig_cc} illustrates some examples of the resulting context condition representations.\vspace{0.2cm}

\noindent \textbf{Step 2: }We can now represent each context situation based on its composing context conditions. This representation can be obtained in two ways: 
\begin{itemize}
	\item \textit{Aggregation:} Each context situation can be represented by a vector with values equal to the mean aggregation of the values of its corresponding composed context condition, as illustrated in Figure~\ref{fig_agr} (also used by~\cite{codina2016distributional}). %(equation \ref{eq_sum}, also used by ~\cite{codina2016distributional}).
	
	\begin{figure}[t]
		\centerline{\includegraphics[width=8cm]{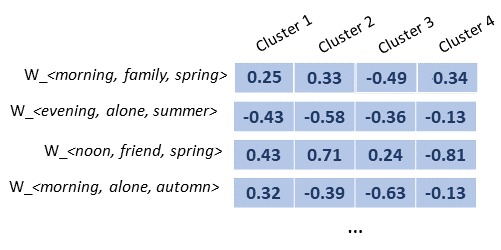}}
		\caption{Examples of representation of context situation by aggregation (Step 2)}
		\label{fig_agr}
	\end{figure}
	\item \textit{Concatenation: } In order to limit the risk of neutralizing the influence of each context condition by the aggregation~\cite{vahidi2018cbpf}, we can represent a context situation with a larger vector built from the concatenation of its composing context conditions (Figure~\ref{fig_concat}).
	
	\begin{figure}[t]
		\centerline{\includegraphics[width=9.5cm]{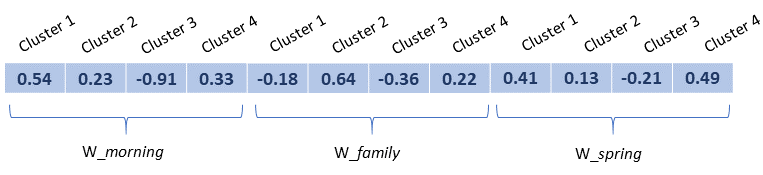}}
		\vspace{-0.3cm}
		\caption{Example of the representation of the context situation $W_{\left\langle morning, family, spring\right\rangle }$ by concatenation (Step 2)}
		\label{fig_concat}
		\vspace{-0.4cm}
	\end{figure}

\end{itemize}

\noindent \textbf{Step 3: } Now, we can find the contexts most similar to the target context situation $s*$ by computing the similarity between every context situation $s$ and the target context situation $s*$, based on the cosine similarity between their vector representations (equation \ref{eq_cosine}, where $d$ is the dimension of the context representation vector $\overrightarrow{w_s}$).

\begin{equation}\label{eq_cosine}
sim(s, s^*)= cosine(\overrightarrow{w_s}, \overrightarrow{w_{s^*}}) = \frac{w_{s}^T w_{s^*}}{\sqrt{\sum_{i=0}^{d}  w_{s,i}^2} \sqrt{\sum_{i=0}^{d}  w_{s^*,i}^2}}
\end{equation}

\noindent \textbf{Step 4: } We select the ratings given in the similar context situations, and make a \textit{local dataset}.\vspace{0.2cm}

\noindent \textbf{Step 5: } We then apply a traditional 2D recommendation technique on this selection of ratings (local dataset), to obtain recommendations (\textit{local model}).%\DC{please put a bibref for the traditional approach}

\section{Experimental Analysis}\label{expSetup}

In this section, we report about our experimental analysis. We first describe the three datasets that we used,  we then  report about parameters used in our approach, and the metrics for experimental evaluation.

\subsection{Datasets}
We evaluated our approach on three real world datasets, which are well-known among the CARS community: (a) \textit{CoMoDa}, a contextual dataset for movie recommendation, collected from surveys~\cite{kovsir2011database}. In this dataset, context situations are defined by 12 different context factors: \textit{time, day type, season, location, weather, social context, end emotion, dominant emotion, mood, physical context, decision} and \textit{interaction},  (b) \textit{STS}~\cite{braunhofer2013context}, a tourism dataset, containing contextual ratings for places of interest, collected using a mobile tourist application. In this dataset, context situations are expressed using 14 context factors: \textit{distance, available time, temperature, crowdedness, knowledge of surroundings, season, budget, day time, weather, companion, mood, weekday, travel goal} and \textit{means of transport}. And (c) the \textit{Music} dataset, contains ratings for contextual music recommendation, collected by an in-car music recommender developed by~\cite{baltrunas2011incarmusic}. In this dataset we have a total number of 8 context factors (\textit{driving style, landscape, mood, natural phenomena, road type, sleepiness, traffic conditions} and \textit{weather}), but it has this specificity that for each context situation, the value of only one context factor is known.\\
Table~\ref{tab:dataset} illustrates some descriptive statistics about these datasets. Note that we calculated the sparsity by means of the following formula: $$1 - \frac{\#ratings}{\#users \times \#items}$$
As Table~\ref{tab:dataset} shows, contrary to the \textit{Music} dataset, the two first datasets are very sparse. In addition, as mentioned before, the \textit{Music} dataset has the disadvantage of a lack of fully context situation information. The ratings of the three datasets go from 1 to 5. But the distribution of the ratings are not similar: in \textit{CoMoDa} and \textit{STS}, the items are mostly well rated, with a mean of around 3.5 and a median of 4. But in the \textit{Music} dataset, the rating distribution is more important in the middle and lower part. In fact we have a median of 2 and a mean of 2.37.

\begin{table}
	\centering
	\caption{Datasets' descriptive statistics}
	\label{tab:dataset}
	\begin{tabular}{lccc}
		\toprule
		Characteristics & \textbf{CoMoDa} & \textbf{STS} & \textbf{Music}\\
		\midrule
		\#ratings & 2296 & 2534 & 4012 \\
		\#users & 121& 325 & 42\\
		\#items & 1197 & 249 & 139 \\
		rating scale & 1-5 & 1-5 & 1-5\\
		rating's mean & 3.83 & 3.47 & 2.37\\
		rating's median & 4 & 4 & 2\\
		rating's standard deviation & 1.05 & 1.29 & 1.48 \\
		sparsity & 98.41\% & 96.86\% & 31.27\% \\
		\#context factors & 12 & 14 & 8\\
		\#context conditions & 49 & 59 & 26\\
		\#items characteristics & 7 & 1 & 2\\
		\#users characteristics & 2 & 7 & 0 \\
		\bottomrule
	\end{tabular}
\end{table}

\subsection{Modeling Parameters}
In this section, we will explain some modeling parameters such as the clustering process, the context similarity threshold and the context-free algorithm used in our experimentations, in order to ensure the reproducibility.

The item/user clustering in the first step needs some pre-treatments:\\
Firstly we have put aside non-characteristic parameters of items/users, like actors and directors in the \textit{CoMoDa} dataset, and artist in the \textit{Music} dataset. By non-characteristic, we mean that they will not be of help for clustering, as each one has a huge number of possible values in comparison to the total number of items/users.\\
Generally, items' or users' characteristics are a mixture of numerical and nominal variables. We have made these uniform by transforming numerical variables such as year and budget of movies in \textit{CoMoDa}, and age of users in \textit{CoMoDa} and \textit{STS} datasets (note that we transform birthday to age in the \textit{STS} dataset, for an easier treatment). 
For the \textit{year} variable we have created two classes: \textit{ancient movies}, those realized before 1988, and \textit{recent movies}, realized after this date. For the \textit{budget} variable, we have made tree segmentations of \textit{weak budget} (less than 18,000,000 \$), \textit{moderate budget} (between 18,000,000 and 50,000,000 \$) and \textit{large budget} (more than 50,000,000 \$). And for \textit{age}, we grouped by interval of 5 years.\\ %\EN{Pourquoi ces choix?}
In some cases where we do not have equally distributed values of variables, a grouping is needed: in \textit{CoMoDa}, for movie language, we have 28 different languages, but 88.61\% of movies are in English. So we have replaced the values of languages other than English by a new value \textit{"other"}, and we have done a similar treatment for the movie country variable; in \textit{STS}, the Point Of Interest (POI) category is defined by a number from 1 to 29. We have kept the POI categories 1, 3, 4 and 9, and we have grouped all others in a single cluster, because the frequency of each one was less than 5\% of the total.

After these pre-treatments, we can cluster items or users. Depending on the available information about items' or users' characteristics, two strategies exist for the clustering:

\begin{itemize}
	\item in case of more than one available characteristic, we can apply a standard clustering algorithm like HC (Hierarchical Clustering) based on these characteristics. We choose the Hierarchical Clustering (HC) technique, which contrary to k-means does not require a pre-defined number of clusters. HC uses a bottom-up approach, it starts to place each item in a cluster, and iteratively merges the two closest clusters, until all the items are merged into a single cluster.
	\item otherwise, if we have only one characteristic (e.g. POI category in \textit{STS} or music category in \textit{Music} datasets), we can directly use this variable as cluster identifier.
\end{itemize} 
So we applied HC on the items and users of \textit{CoMoDa}. As result, the best segmentation proposed was 4 items' clusters and 5 users' clusters.
In the case of \textit{STS}, we obtain 2 clusters of users by HC. But for clustering items, as we had only one characteristic about them, which is the POI category, we used it directly as cluster number. And finally for \textit{Music}, where we had 2 characteristics about items, the artist name and the music category, we used the music category, which is between 1 and 10, as items cluster number.\\
In step 3 we need to set a similarity threshold for identifying the most similar contexts. Empirically, we set this threshold equal to 0.5, which means that when we want to select local datasets, we select ratings that have been given in context situations which are more than 50\% similar to the target user's context situation.

The traditional (context-free) recommendation technique used in the last step of our approach is the Biased Matrix Factorization model~\cite{koren2009matrix}, which is one of the best-performing techniques reported in the state of the art~\cite{codina2016distributional} (from LibRec Java API \cite{guo2015librec}).

\subsection{Evaluation Parameters}
For the evaluation of our approach, we avoided to exclude items or users with low counts, in order to match as closely as possible the conditions of real recommendation applications. Due to the relatively small size of our dataset, we evaluated our approach based on 5-fold cross-validation. As many researches in the domain, we used MAE (Mean Absolute Error) and RMSE (Root Mean Squared Error) metrics to evaluate the rating estimation. These metrics compute the difference between the actual and predicted ratings, but the RMSE penalizes large errors more. Lower values of these metrics show better performances. Note that, because of the relatively small size of the datasets, we didn't use metrics such as precision and recall, where we will obtain very low values.

\section{Results and Discussion}\label{results}

We evaluated our approach in 3 steps: (a) we compared the performances of the derived versions of our approach with each other, (b) we compared our context-aware recommendation approach with a context-free recommendation approach and two baselines, and (c) we compared our approach with similar state of the art CARS approaches to ours.\vspace{0.2cm}

\begin{table}
	\centering
	\caption{MAE/RMSE of the derived techniques of our CBPF approach}
	\label{tab:itemUser}
	\begin{tabular}{lcccccc}
		\toprule
		%\multicolumn{2}{cc}&{CoMoDa}\\
		& \multicolumn{2}{c}{\textbf{CoMoDa}} & \multicolumn{2}{c}{\textbf{STS}} & \multicolumn{2}{c}{\textbf{Music}}\\
		Models&MAE&RMSE&MAE&RMSE&MAE&RMSE\\
		\midrule
		CBPF-IB & 0.84 & 1.055& 0.95 & 1.19 & 1.25 & 1.50 \\
		CBPF-CIB-AG & 0.82 & 1.03 & 0.84 & 1.08 & 1.05 & 1.29 \\
		CBPF-CIB-CN & \textbf{0.73} & \textbf{0.93} & 0.82 & 1.03 & \textbf{0.72} & \textbf{0.87}\\
		\bottomrule
		CBPF-UB & 0.85 & 1.06 & 0.96 & 1.20 & 1.06 & 1.30\\
		CBPF-CUB-AG & 0.83 & 1.04& 0.85 & 1.08 & --- & --- \\
		CBPF-CUB-CN & 0.81 & 1.02& \textbf{0.80} & \textbf{1.02} & --- & ---\\
		\bottomrule
	\end{tabular}
\end{table}

Table \ref{tab:itemUser} illustrates the performances of the derived techniques of our approach, in terms of rating estimation. \textit{CBPF-IB} and \textit{CBPF-UB} refer to the item- and user-based correlation model, \textit{CBPF-CIB-AG} and \textit{CBPF-CUB-AG} refer to the correlation models based on the cluster of items or users, with the aggregation technique, and finally \textit{CBPF-CIB-CN} and \textit{CBPF-CUB-CN} refer to the same model, but with the concatenation technique.\\  
As expected, the last version, which is the cluster-based approach with the concatenation technique for the context representation (\textit{CBPF-CIB-CN} and \textit{CBPF-CUB-CN}), has the best performances. In fact by clustering items/users we not only gain in term of computation cost but also in term of performance. We can explain this gained performance by the fact that in general, a correlation computation gives more precise results with a larger number of data. And the clustering allows to gather more data together, and so results on a better correlation computation and global performance. Moreover, the concatenation technique allows to preserve the real influence of each one of the context conditions, and so to gain in performance. \\
Another interesting point is that there is not a single winner between the item-based influence model and the user-based. As we can see in the table, contrary to \textit{STS}, where \textit{CBPF-CUB-CN} has the best performance, in \textit{CoModa}, the item-based model gives better results (with \textit{CBPF-CIB-CN}). So we can say that the choice between the item- or user-based model depends on the data, and in particular on  the  amount of available information about items'/users' characteristics. 
%In particular,  results in Table \ref{tab:itemUser} depend on the  amount of available information about items'/users' characteristics. 
In fact, as Table~\ref{tab:dataset} shows, for \textit{CoMoDa}, we have more characteristics about items than about users, while the opposite holds for \textit{STS}. So we can say that  having  more items' or users' characteristics implies better clusters. %\DC{I found the previous sentence a bit unclear, reformulated, pls check}
And as a result, we will be able to compute more relevant correlations and create a better model to do the recommendations.
Note that we couldn't compare this effect on the \textit{Music} dataset, because we did not have information about users' characteristics.\\

\begin{figure}[ht]
	\centering
	\begin{tikzpicture}
	\begin{axis}[
	ybar,
	%area legend,
	bar width=9.5,
	width=7cm,
	enlargelimits=0.30,
	legend style={at={(0.5,-0.15)},
		anchor=north,legend columns=-1, font=\small},
	ylabel={\#MAE Improvement (\%)},
	symbolic x coords={CoMoDa, STS, Music},
	xtick=data,
	nodes near coords,
	nodes near coords align={vertical},
	]
	\addplot coordinates {
		(CoMoDa,15.9) (STS, 19.3) (Music, 27.6) 
	};
	\addplot coordinates {
		(CoMoDa,14.3) (STS, 15.9) (Music, 38.8) 
	};
	\addplot coordinates {
		(CoMoDa,12.1) (STS,28.3) (Music, 43.2) 
	};
	\legend{MF, Exact Pre-filtering, Binary Pre-filtering}
	\end{axis}
	\end{tikzpicture}
	\caption{MAE improvement (\%) of our approach with respect to context-free MF and baselines}
	\label{MAEreduction}
\end{figure}
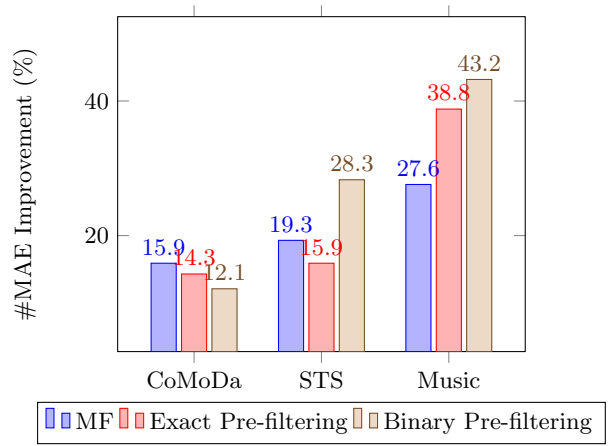
%Les valeurs de cet figure doivent etre vérifiée!

Figure~\ref{MAEreduction} illustrates the MAE improvement that our approach makes over the context-free recommendation and the baselines. The context-free recommendation technique used in this experimentation is a Matrix Factorization (MF) technique named BiasedMF~\cite{koren2009matrix}. The comparison of our context-aware recommendation and this context-free matrix factorization confirms that users' contextual information can help the recommender to improve its performance.\\
As baselines, we used the \textit{exact pre-filtering} approach~\cite{adomavicius2005incorporating} and a second baseline (\textit{binary pre-filtering}) that we built as follow: we represented the context by means of a binary vector with a size equal to the total number of context conditions, where the value of each cell is equal to 1 if the corresponding context condition is present in the context situation, or equal to 0 if it is not present. We did a pre-filtering recommendation using this binary context representation.
As the Figure~\ref{MAEreduction} shows, our approach outperforms these two baselines.
The improvement over the \textit{exact pre-filtering} shows that the idea of filtering the ratings based on the ones done in similar contexts is effective. And the improvement over the \textit{binary pre-filtering} shows the positive effect of representing the context based on the influence of context on ratings.

%\begin{figure}
%	\centering
%\begin{tikzpicture}
%\begin{axis}[
%ybar,
%bar width=7,
%width=9cm,
%enlargelimits=0.15,
%legend style={at={(0.5,-0.30)},
%	anchor=north,legend columns=-1},
%ylabel={\#Rating estimation Error},
%symbolic x coords={MF,Exact Pre-filtering, CBPF-IB, CBPF-CIB-AG, CBPF-CIB-CN, CBPF-UB, CBPF-CUB-AG, CBPF-CUB-CN},
%xtick=data,
%nodes near coords,
%nodes near coords align={vertical},
%x tick label style={rotate=45,anchor=east},
%]
%\addplot coordinates {
%	(MF, 1.0) (Exact Pre-filtering, 1.0) (CBPF-IB, 0.84) (CBPF-CIB-AG,0.82) (CBPF-CIB-CN,0.73) (CBPF-UB,0.85) (CBPF-CUB-AG,0.83) (CBPF-CUB-CN,0.81)
%};
%\addplot coordinates {
%	(MF, 1.0) (Exact Pre-filtering, 1.0) (CBPF-IB, 1.05) (CBPF-CIB-AG,1.03) (CBPF-CIB-CN,0.93) (CBPF-UB,1.06) (CBPF-CUB-AG,1.04) (CBPF-CUB-CN,1.02)
%};
%\legend{MAE,RMSE}
%\end{axis}
%\end{tikzpicture}
%\caption{Rating estimation of \textit{CoMoDa} dataset}
%\end{figure}

\begin{table}
	\centering
	\caption{Comparison with state of the art}
	\label{tab:statOfTheArt}
	\begin{tabular}{lcccccc}
		\toprule
		%\multicolumn{2}{cc}&{CoMoDa}\\
		& \multicolumn{2}{c}{\textbf{CoMoDa}} & \multicolumn{2}{c}{\textbf{STS}} & \multicolumn{2}{c}{\textbf{Music}}\\
		Models&MAE&RMSE&MAE&RMSE&MAE&RMSE\\
		\midrule
		CBPF & \textbf{0.73} & \textbf{0.93} & \textbf{0.80} & \textbf{1.02} & \textbf{0.72} & \textbf{0.87} \\
		DSPF & 0.86 & 1.08 & 1.26 & 1.62 & 1.76 & 2.49 \\
		Deviation-based CAMF & 0.76 & 1.02 & 1.03 & 1.37 & 0.82 & 1.06\\
		Similarity-based CAMF &  \textbf{0.73} & \textbf{0.92} & 0.94 & 1.17 & \textbf{0.72} & 1.09é\\
		DCM & 0.79 & 1.04 & 0.96 & 1.24 & 1.11 & 1.41 \\
		\bottomrule
	\end{tabular}
\end{table}

Finally, we compared our approach with four state of the art approaches, which are more closer to our approach: (a) \textit{DSPF} (Distributional Semantic Pre-Filtering) \cite{codina2016distributional}, (b) \textit{Deviation-based CAMF} (Context-Aware Matrix Factorization) \cite{baltrunas2011matrix}, (c) \textit{Similarity-based CAMF}~\cite{zheng2015similarity} and (d) \textit{DCM} (Differential Context Modeling)~\cite{zheng2012optimal}. In fact \textit{DSPF} and \textit{deviation-based CAMF} approaches try to model the context based on the influence of contexts on ratings, and \textit{similarity-based CAMF}, \textit{DCM} and \textit{DSPF} uses the similarities among contexts in their approaches. Each one of these approaches have different versions (cited in Section~\ref{relatedWork}). We tested all the possible versions, and the performances of the best version of each approach, in terms of rating estimation are illustrated in Table~\ref{tab:statOfTheArt} for each dataset. We tested the state of the art algorithms by relying on the CARSkit Java API \cite{zheng2015carskit}.  So for the \textit{CoMoDa} dataset, we report the performances of CBPF-CI-CN, DSPF-IB, CAMF-CU, CAMF-ICS and DCW. For \textit{STS}: CBPF-CU-CN, DSPF-IB, CAMF-CU, CAMF-ICS and DCW, and for \textit{Music}: CBPF-CI-CN, DSPF-UB, CAMF-CU, CAMF-ICS and DCW.\\
The reported results are average of multiple executions based on 5-fold cross-validation. For each dataset, the values in bold are statistically significant better (95\% confidence level) than other approaches. The statistical significance has been calculated using the Wilcoxon rank test.\\
The illustrated performances in Table~\ref{tab:statOfTheArt} show that, for all the three datasets, our approach can have better or comparable performances in comparison to state of the art (lower values of these metrics show better performances). Note that when comparing to CAMF, we obtain better, even if comparable, performances. However, our pre-filtering approach has the important advantage of being easily pluggable into any existing  recommender system which is already in production, in order to improve it, while adopting CAMF would imply to re-implement the whole process.

Finally, we can state that our proposed modeling approach is able to produce good performances on roughly different datasets from various domains. In fact, the three datasets used in our experimentations are from three different domains (movie, tourism and music), with different characteristics in terms of density, rating distribution and the number of available context and content information.
%\EN{Indeed, the three datasets tested are of different densities (31.27\% for \textit{Music} and greater than 96\% for \textit{STS} and \textit{Comoda}), which indicates that our approach is able to give good results whatever the density of the initial utility matrix ($Users \times Items \times Context \rightarrow r$). In addition, the number of items (rather low for \textit{Music} - $139$ - and stronger for \textit{Comoda} - $1197$) and/or the number of users (rather low for \textit{Music} - $42$ - and relatively high for \textit{STS} - $325$), do not seem to impact the performance of our approach. Likewise, neither the number of context conditions (rather weak for \textit{Music} - $26$ - and rather strong for \textit{STS} - $59$), nor the number of item characteristics (less than 2 for \textit{STS} and \textit{Music}) and/or the number of user characteristics (less than 2 for \textit{Music} and \textit{Comoda}) do not prevent our approach to perform.} 
Moreover, clustering items or users has shown to have beneficial effect. In fact, our results indicate that grouping items or users can roughly help the model to catch more significantly the influence of context on ratings. 

\section{Conclusions and Further Work}

%\DC{we should not insist on 'our previous ws...', it is sufficient to put a statement in the intro} \ZVF{I agree. Rephrased. }\DC{As we have already outlined in the intro that this paper extends, I propose to put 'In this paper, we present a correlation-based pre-filtering (CBPF) approach,....'}
In this paper, we present a correlation-based pre-filtering (CBPF) approach, that integrates contextual information into the recommendation process by modeling the influence of the context on ratings. This influence is computed by the item- or user-based Pearson Correlation Coefficient (PCC) between the context and the ratings. CPBF tries to take advantage of content and context information in its recommendation process to improve its performances.
%More precisely, CPBF relies on five steps. During the first step, the influence of the context conditions on the ratings is represented using the Pearson Correlation Coefficient. This correlation can be item-based or user-based. Note also that to overcome computational costs due to the large number of items/users, we cluster items/users based on their characteristics. During the second step, context conditions representations are aggregated using \ZVF{average} or concatenation to represent context situations. During the third step, context situations similar to the current context situation are detected. The corresponding ratings are collected in a \textit{local dataset} during step four. Finally, a traditional 2D recommender system is applied on this \textit{local dataset}.\\ 
In our experimental analysis, we evaluate our approach on three different contextual datasets (\textit{Comoda}, \textit{STS} and \textit{Music}), and analyze  performances  based on the characteristics of these datasets. Experiments validate the positive effect of taking into account contextual information about the user in the recommendation process, and show that our approach outperforms state of the art techniques  in most cases. Furthermore, experimental results show that the PCC can efficiently catch the influence of context on ratings. Also, due to the large number of items/users, grouping them not only reduces computational cost, but also increases performances.\\

%\DC{to me there is too much detail in this conclusion, especially concerning the 5-steps, do we really need to re-mention/describe them? } \ZVF{I agree too. Rephrased.}
%\DC{rephrased a bit above, but where we put results about computational cost?}

As just mentioned, by clustering items/users, we gain in terms of computation cost. We can gain even more by clustering context situations. So in the future, we would like to apply a clustering on context situations (like \cite{codina2016distributional}) to limit the local model building computations. \\
In our approach, we compute the influence of context on ratings based on the PCC. In the future we plan to test other statistical models (e.g., ANOVA) for the correlation computation.\\
In our experiments we cluster items/users based on their available characteristics information. But this kind of information is not always available, so it would be interesting to test other clustering strategies, like clustering items/users based on ratings.\\ 
In real world applications, not all context factors have the same importance and impact on ratings. Depending on the application, some context factors can play a more important role than others. For example, in the case of recipe recommendation, factors like season, available tools around the user, and her cooking competence would
be more important. While in music recommendation, activity
and psychological context would be more influencing. So in future work we plan  to take this fact into consideration, and rely on weighted context factors, based on their importance.

\bibliographystyle{splncs04}
\bibliography{biblio}

\end{document}